\documentstyle[preprint,aps]{revtex}
\begin{document}

%
%

\title{  Hole Pockets in the Doped 2D Hubbard Model}

\author{ Adriana Moreo and Daniel Duffy}

\address{Department of Physics, National High Magnetic Field Lab and MARTECH,
Florida State University, Tallahassee, FL 32306, USA}

\date{\today}
\maketitle

\begin{abstract}

The electronic momentum distribution ${\rm n({\bf k})}$
of the two dimensional Hubbard model is
studied for different values of the coupling ${\rm U/t}$, electronic density
${\rm \langle n \rangle}$, and
temperature, using quantum Monte Carlo techniques.
A detailed analysis of the data on $8\times 8$ clusters
shows that features consistent
with hole pockets at momenta ${\rm {\bf k}=(\pm {\pi\over{2}},\pm
{\pi\over{2}})}$
appear as the system is doped away from half-filling. Our results are
consistent with recent experimental data for the cuprates discussed
by Aebi et al. (Phys. Rev. Lett. {\bf 72}, 2757 (1994)).
In the range of couplings studied, the depth of the pockets is maximum at
${\rm \langle n \rangle \approx 0.9}$, and it increases with decreasing
temperature. The apparent absence of hole pockets in
previous numerical studies of this model is explained.

\end{abstract}

\pacs{74.20.Mn, 74.25.Ha, 74.25.Jb}

%
%
One of the simplest models often used to describe the physics of the
high temperature cuprate superconductors is defined by
the two dimensional (2D) Hubbard Hamiltonian.
Several normal state properties of these materials,
such as antiferromagnetism,
the behavior of the magnetic susceptibility
with temperature and density, the optical conductivity
and others, are
qualitatively reproduced by this model.\cite{review}
 Although superconductivity has not been
observed in Monte Carlo simulations\cite{pairing},
 self-consistent techniques suggest
its existence at low temperatures in the ${\rm d_{x^2 - y^2}}$
channel.\cite{pao}
In order to understand the pairing mechanism leading to such
superconducting
instability it is important to study the shape of the Fermi
surface in this model, and compare it against angle-resolved
photoemission experiments
for the cuprates. Theories based on antiferromagnetism (AF) explain in a
natural way the apparent d-wave symmetry of the condensate, but they also
suggest the existence of ``hole pockets'' in ${\rm n({\bf{k} })}$.\cite{Bob}
Since for some time evidence of a large electron-like Fermi
surface has been reported in experiments\cite{nopock}, ideas based on AF
have been questioned.
Moreover, in agreement with these experiments, theoretical studies
of  the Hubbard\cite{Hub} and t-J Hamiltonians\cite{Steph},
reported electron-like Fermi surfaces very similar in shape to the
non-interacting case, even at low hole densities where antiferromagnetic
correlations should still be strong in the system. Possible many-body
effects have been invoked to explain such a behavior.

However,
on the experimental side, these results have been recently challenged
by Aebi et al.\cite{aebi} Using a photoemission technique that allows
the mapping of the whole Fermi surface, these authors have shown
the existence of hole pockets in their data for Bi2212 consistent with
a rigid band filling of the hole states predicted by
antiferromagnetically based mean-field approximations.\cite{Bob}
In parallel to the experimental results,
recent theoretical arguments by Dagotto et al.\cite{Elb} conjectured
that the large Fermi surface of the 2D Hubbard model observed
at finite temperature can be made compatible with the existence of
hole pockets. Their reasoning is that unless the temperature of the
simulation is smaller than the energy difference of holes states at
momenta ${\rm {\bf k} = (\pm {\pi\over{2}}, \pm {\pi\over{2}}) }$ and
$(0,\pi),(\pi,0)$
(which they found to be very small
 at ${\rm U/t=10}$), the hole pocket effect is washed out
for thermodynamical reasons. These authors suggested that
studies at stronger couplings, lower temperatures and larger lattices
should provide evidence for the existence of hole pockets.

Motivated by these challenging experimental and theoretical ideas,
in this paper a  careful analysis of Quantum Monte Carlo data for the 2D
Hubbard model is discussed.
We conclude that numerical evidence indeed
supports the existence of hole pockets at ${\rm {\bf k}=(\pm {\pi\over{2}},\pm
{\pi\over{2}})}$
at low hole density. Thus, our results reopen the possibility
that the normal state properties of the
high temperature superconductors at half-filling can be qualitatively
approximated by
a band filling of the states obtained in the mean-field
spin-density-wave
approximation. The Hubbard Hamiltonian is given by
$$
{\rm H=
-t\sum_{<{\bf{ij}}>,\sigma}(c^{\dagger}_{{\bf{i}},\sigma}
c_{{\bf{j}},\sigma}+h.c.)+
U\sum_{{\bf{i}}}(n_{{\bf{i}} \uparrow}-1/2)( n_{{\bf{i}}
\downarrow}-1/2)+\mu\sum_{{\bf{i}},\sigma}n_{{\bf{i}},\sigma} },
\eqno(1)
$$
\noindent where ${\rm c^{\dagger}_{{\bf{i}},\sigma} }$ creates an electron at
site ${\rm {\bf i } }$
with spin projection $\sigma$, ${\rm n_{{\bf{i}},\sigma} }$ is the number
operator, and the sum
${\rm \langle {\bf{ij}} \rangle }$
runs over pairs of  nearest neighbor lattice sites. ${\rm U}$ is the
on site coulombic repulsion, ${\rm t}$ the hopping amplitude, and $\mu$ is the
chemical potential. This Hamiltonian will be analyzed using standard
Quantum Monte
Carlo techniques.\cite{Blan} Clusters with up to $8 \times 8$ sites will be
studied varying the
temperature ${\rm T}$, filling ${\rm \langle n \rangle }$,
and coupling ${\rm U/t}$. The notorious sign-problem  prevents us
from working at very low temperatures and large couplings. Those used in
this paper correspond to the lowest T and largest U/t that can be
reliably reached with confidence using the QMC algorithm.\cite{param}
We will show
below that they are enough to observe hole pockets in the Hubbard model.

The first issue to analyze is whether
antiferromagnetic correlations are strong in the system away from
half-filling, since this is a necessary condition for the existence of
hole pockets induced by AF.
In Fig.1-a,
the spin-spin correlation function ${\rm C({\bf r})=(-1)^{|r|} \langle
S^z_{\bf i} S^z_{\bf i+r} \rangle }$ is presented for ${\rm U/t=6}$ on an
$8\times 8$ lattice
at ${\rm T=t/4}$. At half-filling, ${\rm \langle
n \rangle =1}$, (filled circles) the correlations are strong and remain finite
even at the largest possible distance clearly indicating the expected
antiferromagnetic long range order (LRO) in the system. More interesting for
our
analysis is that at density ${\rm \langle n \rangle =0.9}$ (open circles)
where antiferromagnetic LRO does not exist,
it is clear that the spin
correlations are strong up to a distance of about 3 to 4 lattice spacings.
Increasing further the hole density to
${\rm \langle n \rangle =0.75}$
(triangles), antiferromagnetism is further suppressed
but it remains strong between nearest-neighbors spin showing that the
moments are still well-formed even at this fairly large hole doping.
When the coupling is reduced to the value that has been more widely
analyzed  by QMC techniques namely ${\rm U/t=4}$,
a qualitatively similar behavior is found at the same temperature.
The spin correlations at this coupling (shown in
Fig.1-b) are less developed, but LRO is still
observed at half-filling. Working at ${\rm \langle n \rangle =0.9}$
short range
antiferromagnetism is also robust as it occurs at higher couplings.
Decreasing the temperature to ${\rm T=t/6}$ (shown in
Fig.1-c) the spin correlations at ${\rm U/t=4}$ become as strong as for
${\rm U/t=6}$ at ${\rm T=t/4}$. Then, Figs.1-a,b,c have shown
that at a realistic density of  ${\rm \langle n
\rangle =0.9}$,
the antiferromagnetic correlations are sufficiently developed that they may
originate hole-pockets in the Fermi surface. Eventually, these pockets should
evolve
into a noninteracting-like Fermi surface as the electronic density is
further decreased from half-filling.

Since the antiferromagnetic correlations are robust,
then the natural question to discuss is why
hole pockets have not been observed in previous studies (which were only
carried out at ${\rm U/t=4}$). To address
this question let us concentrate on the analysis of the momentum distribution
 ${\rm n({\bf{k}})=\sum_{\sigma}
c^{\dagger}_{{\bf{k}},\sigma}c_{{\bf{k}},\sigma} }$, where
${\rm c^{\dagger}_{{\bf{k}},\sigma}=\sum_{{\bf{j}}}e^{i\bf{k.j}}
c^{\dagger}_{{\bf{j}},\sigma} }$. Note that
${\rm n({\bf{k}}) }$ has been previously
measured\cite{Hub,Steph} but in these and some other papers, the shape of
the Fermi surface was determined by calculating the position of the
momenta where ${\rm n({\bf{k}})=0.5}$. In principle, this procedure is correct.
However, using such a convention here we show that the pockets could be
overlooked due to finite temperature and/or small lattice effects.
To understand
this point, in Fig.2-a ${\rm n({\bf k }) }$ versus ${\rm {\bf k}}$ is shown
along the
${\rm k_x=k_y}$ direction for ${\rm U/t=6}$ at ${\rm T=t/4}$
on an $8\times 8$ cluster and density
${\rm \langle n \rangle =0.9}$ (open squares). Comparing the interacting
results
with noninteracting ${\rm U/t=0}$ data for the same parameters (filled
squares) we observe that,
in both cases, ${\rm n({\bf{k}})}$ becomes ${\rm 0.5}$
at approximately the same momentum. Repeating the calculation along other
directions in momentum space, and $if$ the ${\rm n({\bf{k}})} \approx
0.5$ criterion is still used, it is thus understandable why a Fermi surface
resembling a
non-interacting system was obtained in previous studies.
However, this does not rule out the existence of pockets.
The crosses shown in Fig.2-a correspond to spin density wave mean-field
(SDW-MF)
results.\cite{Bob} In this approximation, an antiferromagnetic state is used
which
effectively produces a $2\times2$ unit cell. The mean-field Hamiltonian
is diagonalized producing conduction and valence bands separated by the
antiferromagnetic gap. The energy levels are given by
$E_k=\pm\sqrt{\epsilon_k^2+\Delta^2}$, where
$\epsilon_k=-2t(cos k_x+cos k_y)$,
and $\Delta$ is found using a self-consistent equation.
At half-filling, the valence band is filled, and
${\rm n(k)}={1\over{2}}(1-{\epsilon_k\over{E_k}})$ is in very good
agreement with the numerical data\cite{foot2} as it is shown in Fig.2-b.
Now, let us assume that a robust
antiferromagnetism survives the introduction of hole doping, as it was
suggested by the results of Fig.1-a,b,c.
Quasiparticles are removed from the top of the valence band to mimic the
presence of doping. ${\rm n({\bf k})}$ now is given by,

$$
{\rm
n({\bf k})={1\over{2}}(1-{\epsilon_{\bf k}\over{E_{\bf k}}})
({1\over{e^{-\beta(|E_{\bf k}|-\mu)}}+1} }),
\eqno(2)
$$

\noindent where $\mu$ is selected such that the
density is ${\rm \langle n \rangle }$.
In Fig.2a-b the SDW mean-field results are compared against the
Monte Carlo data. The agreement along the ${\rm k_x=k_y}$ direction
both at half-filling and at finite hole density is
remarkable. We have explicitly checked that only at densities where the
antiferromagnetic
correlations become of one lattice spacing or less (${\rm \langle n \rangle
\approx 0.75}$), the comparison between the numerical data and the SDW
results deteriorates. Reducing further the density, the QMC data converges
smoothly to a weakly interacting gas of electrons.

The agreement between a theory based on strong AF correlations and the
numerical data is consistent with the strong AF correlations
shown in Fig.1. However, the main issue addressed
in this paper remains paradoxical, i.e. if the numerical data and the
SDW
approximation are in good agreement, why there is no hole pocket in the
occupation number mean value of Fig.2-a?
To understand this point, it is instructive to study the effects that high
temperatures and a finite lattice produce in ${\rm n({\bf k})}$ in the
mean-field approximation. In Fig.3-a we show the mean field ${\rm n({\bf k})}$
along the
${\rm k=k_x=k_y}$ direction on a $20\times20$
lattice, for ${\rm U/t=6}$ and ${\rm \langle n \rangle =0.9}$. At ${\rm T=0}$
there is a clear pocket-like feature at $({\pi\over{2}},{\pi\over{2}})$, very
different
from that observed numerically in Fig.2a.
However, as the temperature increases the pocket becomes less pronounced
and for temperatures ${\rm T=t/6}$ and ${\rm t/4}$, which are the
values most commonly studied with Monte Carlo techniques,
it has all but disappeared. In
Fig.2-a the crosses are the mean-field results and it is clear that they
are in much better agreement with the Monte Carlo data than the ${\rm U/t=0}$
points. This shows that the existence of pockets along the ${\rm k=k_x=k_y}$
direction is a feature very difficult to see and quickly washed out by high
temperature effects\cite{foot} due to the rapid change in ${\rm n({\bf k})}$
where
the pockets are expected. In Fig.3-b, we show the region on
the ${\rm U/t}$-${\rm T/t}$ plane
where a pocket, defined as a local $minimum$ in the momentum distribution,
is expected at $({\pi\over{2}}, {\pi\over{2}})$
according to the SDW mean-field results at ${\rm \langle n \rangle =0.9}$.
In the same plot the crosses indicate the lowest
temperatures that can be presently
reached for each ${\rm U/t}$
value with the QMC algorithm, the limitation being caused mainly by the
well-known sign-problem. Clearly, all
the crosses lie in a region where pockets are not observed in the
mean-field approximation. Thus, the absence of a local minimum in our results
of
Fig.2-a are not incompatible with hole pockets, and it is caused by
finite temperature and small lattice effects.

The most striking evidence showing that the interacting
system is different from the non-interacting one is obtained
analyzing the line from ${\rm {\bf k} = (0,\pi)}$ to ${\rm (\pi,0)}$
(i.e. from Y to X).
The SDW-MF predicts a constant momentum distribution  along $cosk_x
+ cosk_y = 0$ even at finite
coupling, but this is a spurious degeneracy of this simple
approximation and thus it is not useful to contrast against the QMC data.
The noninteracting system ${\rm U/t=0}$, also has ${\rm n({\bf k})}$ constant
along
this line at any density and temperature, and it is given by
$$
{\rm n({\bf k})={1\over{e^{\beta(\epsilon_{\bf
k}-\mu)}+1}}={1\over{e^{-\beta\mu}+1}}.}
\eqno(3)
$$
\noindent
At half-filling $\mu=0$ and ${\rm n({\bf k})}$ along this line is
0.5. This symmetry allows us to study pockets in the interacting
system by looking for a reduction along the X-Y line of the momentum
distribution as the coupling is increased. If ${\rm
n({\pi\over{2}},{\pi\over{2}})}$ is
the smallest along this line, it would be an evidence in favor of hole pockets.
In Fig.4-a ${\rm n({\bf k})}$ vs ${\bf k}$
along the $(\pi,0)-(0,\pi)$ line
is presented, using an $8\times 8$ cluster,
${\rm U/t=6}$, and ${\rm T=t/4}$. In the half-filled case (filled
circles) ${\rm n({\bf k})}$ is
constant even for an interacting system.
However, at density ${\rm \langle n \rangle=0.9}$ (open circles)
there is a clear minimum (pocket) at ${\bf k}=({\pi\over{2}},{\pi\over{2}})$.
The minimum lasts as long as antiferromagnetism is present in the
system, i.e. at density
${\rm \langle n \rangle \leq 0.75}$ or smaller the minima are no longer
observed giving support to our interpretation of this feature as hole pockets
caused by AF correlations.

In Fig.4-b, similar results are presented for ${\rm U/t=4}$. Here the
pockets are smaller than at ${\rm U/t=6}$
because the AF order is not so strong
but the qualitative behavior
is the same. The maximum depth occurs again for ${\rm \langle n \rangle =0.9}$.
In Fig.4-c we show
the pocket at several couplings ${\rm U/t}$
ranging from 0 to 6 at a fixed density ${\rm \langle n \rangle =0.9}$. As the
coupling increases the pocket becomes deeper which is intuitively
understandable using the AF picture.
Finally, in Fig.4-d the
dependence with temperature at ${\rm U/t=4}$ and ${\rm \langle n \rangle
=0.9}$ is presented. As with the coupling dependence,
the pocket becomes deeper as the temperature is lowered from ${\rm T=t/2}$ to
${\rm t/6}$.\cite{u8} It is clear that the behavior of
the pocket is correlated with the behavior of the magnetic
correlations presented in Fig.1. In other words, when the spin correlations are
strong, the pockets appear in the spectra.
When the correlations increase, due to an increase in
coupling or decrease in temperature, the pockets become deeper. But when
the AF correlations are of the order of one lattice spacing,
as for ${\rm \langle n \rangle < 0.75}$, the pockets are no longer
observed, and ${\rm n({\bf k})}$ becomes constant along the $(\pi,0),(0,\pi)$
direction as in the noninteracting case.

In this paper, we have presented evidence that in the 2D Hubbard
model the strong antiferromagnetic fluctuations produce
hole pockets at momenta
${\rm {\bf k}=(\pm {\pi\over{2}},\pm {\pi\over{2}})}$. Monitoring the behavior
of ${\rm n({\bf k})}$ along the line $(\pi,0),(0,\pi)$, where the electronic
momentum distribution is constant in
the noninteracting system, we have observed a minimum at
$({\pi\over{2}},{\pi\over{2}})$ compatible
with the existence of a hole pocket. Along the ${\rm k_x=k_y}$ direction
no local minimum  is numerically observed but  by comparing our results to a
SDW-MF approximation we have shown
that this feature is due to finite temperature effects, as predicted by
Dagotto et al.\cite{Elb}
For the values of ${\rm U/t}$ here
studied, i.e. 4 and 6, the pockets exist for $\langle n \rangle > 0.75$
and the maximum effect is observed at ${\rm \langle n \rangle =0.9}$. Larger
couplings and lower temperatures would certainly increase
the range of existence of these hole pockets. Thus, we have shown that
the recent experiments of Aebi et al. are compatible with results obtained
for the 2D Hubbard model giving support to theoretical descriptions of
the cuprates based on antiferromagnetic correlations.


\medskip
We thank E.~Dagotto
for useful conversations
and suggestions.
A. M. is supported by the Office of Naval Research under
grant ONR N00014-93-0495. We thank SCRI and the Computer Center at FSU for
providing us access to the Cray-YMP and ONR for giving us access to the
CM5 connection machine.

\medskip

\vfil\eject

%
%

{\bf Figure Captions}

\begin{enumerate}

\item
(a) Spin-spin correlation ${\rm C({\bf r})=\langle S^z_{\bf i} S^z_{\bf
i+r}\rangle
(-1)^{|{\bf r}|} }$ for ${\rm U/t=6}$, ${\rm T=t/4}$ on an $8\times 8$ lattice
at
different fillings. The error bars are of the size of the dots.

(b) Same as (a) but for ${\rm U/t=4}$.

(c) Same as (b) but for ${\rm T=t/6}$.

\item
(a) ${\rm n({\bf k})}$ as a function of momentum along the
${\rm k_x=k_y}$ diagonal direction on an
$8\times 8$ lattice for ${\rm U/t=6}$, ${\rm T=t/4}$ at
${\rm \langle n \rangle =0.9}$.
The open squares are Monte Carlo results, the filled squares
are ${\rm U/t=0}$ results, and the crosses correspond to the SDW
mean field approximation (see text).

(b) Same as (a) for ${\rm \langle n \rangle =1.0}$ (half-filling).

\item
(a) Mean field values of ${\rm n({\bf k})}$ vs momenta along the diagonal
${\rm k_x=k_y}$ direction using a $20\times 20$ lattice, ${\rm U/t=6}$,
${\rm \langle n \rangle =0.9}$ for
${\rm T/t=0, 1/50,1/20,1/6}$ and ${\rm 1/4}$.

(b) Mean field determination of the region where pockets, define as a
relative minimum along the ${\rm k_x=k_y}$ direction, are observed in
the ${\rm U/t}$ vs ${\rm T/t}$
plane for ${\rm \langle n \rangle =0.9}$
on a $20 \times 20$ cluster. The
crosses indicate the lowest temperatures that can be reached with our
Monte Carlo technique for different values of the coupling, due to the
sign problem.

\item
(a) Momentum distribution ${\rm n({\bf k})}$ as a function of momenta along the
$(\pi,0)$ to $(0,\pi)$ direction on an $8\times 8$ cluster for ${\rm U/t=6}$,
${\rm T=t/4}$ at ${\rm \langle n \rangle =1}$
(filled circles), ${\rm \langle n \rangle =0.9}$ (open circles) and
${\rm \langle n \rangle =0.75}$ (filled
triangles).

(b) Same as (a) for ${\rm U/t=4}$.

(c) Momentum distribution ${\rm n({\bf k})}$ as a function of momentum along
the
$(\pi,0)$ to $(0,\pi)$ direction on an $8\times 8$ lattice at ${\rm
\langle n \rangle =0.9}$ and
${\rm T/t=1/4}$ for (from above to below) ${\rm U/t=6, 4, 2}$ and ${\rm 0}$.

(d) Momentum distribution ${\rm n({\bf k})}$ as a function of momentum along
the
$(\pi,0)$ to $(0,\pi)$ direction on an $8\times 8$ lattice at ${\rm
\langle n \rangle =0.9}$ and
${\rm U/t=4}$ for temperatures (from above to below) ${\rm T=t/2, t/3,
t/4}$  and ${\rm t/6}$.

\end{enumerate}


\begin{references}
\bibitem{foot2} For values of ${\rm U/t}$ smaller than 6, i.e. for ${\rm
U/t=4}$ ,
the mean field result gives a reasonable qualitative fit to the Monte
Carlo data. Quantitative agreement is obtained by reducing $\Delta$\cite{Hub}.
\bibitem{foot} In small lattices, .i.e. $4\times 4$, $8\times 8$, not
even at ${\rm T=0}$ the pockets
are observed at the
mean field level along the ${\rm k_x=k_y}$ direction.
\bibitem{u8} We also studied couplings ranging from ${\rm U/t=0}$ to 8 at
${\rm T=t/3}$ and we observed that at ${\rm \langle n \rangle=0.9}$
the depth of the
pocket increased with ${\rm U/t}$. However, the pocket is deeper for
${\rm U/t=6}$
and ${\rm T=t/4}$ than for ${\rm U/t=8}$ and ${\rm T=t/3}$.
This can be understood
by noting in Fig.2-b that the first case is closer to the pocket region
than the second.
\end{references}
\end{document}